\documentstyle[BIB]{report}    
\renewcommand{\theequation}{\thesection.\arabic{equation}}
\newcommand{\be}{\begin{equation}}
\newcommand{\ee}{\end{equation}}
\def\lsim{\raise0.3ex\hbox{$<$\kern-0.75em\raise-1.1ex\hbox{$\sim$}}}
\def\gsim{\raise0.3ex\hbox{$>$\kern-0.75em\raise-1.1ex\hbox{$\sim$}}}

\title{ Multi-quark Energies in SU(2)  Lattice Gauge Theory}
\author{ A.M. Green\thanks{E-mail address:GREEN@PHCU.HELSINKI.FI},\\
 Research Institute for Theoretical Physics,  University
of Helsinki, Finland, \\
 C. Michael\thanks{E-mail address:CM@UKACRL},\\
DAMTP, University of Liverpool, UK, \\
 J.E. Paton\thanks{E-mail address:PATON@PH.OX.AC.UK},\\
Department of  Theoretical Physics, University of Oxford, UK \\
 and M.E. Sainio\thanks{During the academic year 1992-93 at Paul Scherrer
Institute, Villigen and ITP, University of Bern,
 \ \ \ Switzerland; on leave
from Dept. of Theoretical Physics, University of Helsinki}\\
 Research Institute for Theoretical Physics,  University
of Helsinki, Finland}

\begin{document}

\large

\maketitle

\begin{abstract}

Energies of four-quark systems have been extracted in a quenched SU(2)
lattice Monte Carlo calculation for two different geometries, rectangular
and colinear, with $\beta=2.4$ and lattice size $16^3\times 32$.
Also, by going to a lattice $24^3\times 32$ and to $\beta=2.5$,
the effect of the finite lattice size and scaling are checked.
An attempt is made to understand these results in terms of a model based on
interquark two-body potentials but modified very significantly
by a phenomenological gluon-field overlap factor.

\end{abstract}

\newpage

\section{ Introduction}
\setcounter{equation}{0}

\vskip 0.5 cm
In the treatment of many-body systems there has been achieved much
success when assuming the bodies interact mainly in pairs. There are
many examples of this in atomic and nuclear physics with the use
of two-body Coulomb and nucleon-nucleon potentials.
 However, at a more basic level in the hadronic world, it is of interest to
 study the many-body system composed of  quarks interacting
via gluons carrying colour.
In this system, especially in view of the special features of the
colour interaction -- such as confinement -- , the continued use of simply
two-body potentials acting between constituent quarks can be questioned.
Unfortunately, even though
 QCD is now well accepted as the fundamental field theory underlying
hadron interactions, the dynamics of multi-quark systems is still too
difficult a problem to be treated at this fundamental level. As a
step towards tackling this problem,
in a series of recent papers\cite{MGP}-\cite{GMP2}, an attempt
has been made to understand the energies of systems of four infinitely heavy
quarks in terms of models  based on interquark potentials i.e. the quarks
are treated as fixed colour sources, and their kinetic energy neglected -- the
static approximation.
This programme requires knowing the following:
\begin{itemize}
\begin{enumerate}
\item Reliable estimates of the {\em exact} energies $(E_i)$ of four-quark
systems. Here the suffix will be taking on the two values 0 and 1 --
corresponding to the ground states and the first excited states respectively.
\item Reliable interquark two-body potentials $V_{qq}$.
\item A physically motivated and convenient model or prescription for
calculating the energies $(E'_i)$ of a four-quark
system in terms of the $V_{qq}$ -- the hope being that the $E'_i$ are a good
approximation to the exact $E_i$.
\end{enumerate}
\end{itemize}
The calculation of the energies $(E_i)$ of a four-quark system is not easy --
even though the basic equation describing the interaction between quarks
and gluons is governed by the well known QCD lagrangian.
 Traditionally, it is thought that the best way
to estimate the $E_i$ is to use a lattice formulation of this basic
lagrangian. To make this reasonably tractable
with present mainframe computers requires introducing the so-called quenched
approximation, in which the creation and annihilation of $q\bar{q}$
pairs is neglected -- details being given in
refs.\cite{GMP2}-\cite{Hunt}.
Over the years, a few such attempts \cite{Markum,Ohta}
have been made to extract the $E_i$ for four-quark systems. However, the
results have always had sufficiently large error bars that it has not proved
possible to carry out the type of analysis suggested in
refs.\cite{MGP}-\cite{GMP2} i.e. they were unable to distinguish between
theoretical models that differed very considerably.
 By taking advantage of advances in computer
technology and also recent theoretical developments in Monte Carlo lattice
simulations, the situation has now improved greatly. In addition,
considering the gauge group SU(2)
 -- and not SU(3) as in refs.\cite{Markum} and \cite{Ohta} -- further
improves the efficiency of the calculation. This use
of SU(2) means the distinction between quarks and antiquarks, in
general, disappears and so will be ignored in most of this paper -- except,
however, in section 5.
Overall, compared with
refs.\cite{Markum} and \cite{Ohta}, the total efficiency can be increased
 by several orders of magnitude, so that accurate and reliable estimates
of the $E_i$ can now be obtained in the static quenched approximation.
Without this latter approximation, present day computers would be unable to
give such estimates.  It should be added that the use of SU(2)
compared with the more realistic world of SU(3) seems to be
reasonable initially, since it has been demonstrated that
the ratios of physical quantities are very similar in SU(2) and SU(3)
e.g. the ratio $m_{GB}/\sqrt{b_S}$, where $m_{GB}$ is the glueball mass
and $b_S$ is the string tension, is close to 3.5 in both
cases \cite{chrism} -- see also ref.\cite{fing} for similar comparisons
involving $T_c$, the temperature for the deconfinement phase transition.

The second requirement mentioned above is the interquark potential
$V_{qq}(r)$, where $r$ is the distance between the two infinitely heavy
quarks in a colour singlet state.
This potential is quite well known and can be
accurately parametrized by essentially the form
\be
\label{Vqqr}
V_{qq}(r)=-\frac{e}{r}+b_Sr.
\ee
The parameters $e$ and $b_S$ can be extracted by fitting the observed masses
of heavy mesons or from Monte Carlo lattice simulations. Since the
four-quark energies ($E_i$) are calculated on a lattice, for consistency
 the $V_{qq}$ used are those calculated as a by-product of the {\em same}
lattice calculation. This improves the accuracy of the predictions, since the
main quantities of interest -- the $E_i$ -- are the {\em differences}  between
the total four-quark energies and that of two separated two-quark systems.
In other words, the $E_i$ can be thought of as the binding energy between
these two two-quark systems.

The final requirement is a model for generating four-quark energies $(E'_i)$
 in terms of the $V_{qq}$. Since the purpose of this work is to demonstrate
the relationship between $E_i$ and $E'_i$, it is appropriate to consider
first only very simple four-quark geometries.
If, in these extreme cases, a model does
not work (i.e. $E'_i\not =E_i$), then there is no reason to believe it will
succeed in more complicated cases. On the contrary, even if the model does work
in
these simple cases, then there is no guarantee of success elsewhere as the
situations become more complicated.
With this philosophy in mind, the first problems
to be treated will involve four infinitely-heavy quarks fixed into some
simple geometry.  This restricted model is consistent
with the corresponding Monte Carlo lattice simulation generating the $E_i$,
since the latter exploits the so-called static quenched approximation.
The restriction to ``simple four-quark
geometries'' is purely for convenience. In fact, in principle, any four-quark
geometry can be treated by the present methods. However, part of the idea
is to start with simple geometries -- such as rectangles or colinear -- in
order to extract systematics. The latter systematics are then parametrized
 in some simple way, with the hope that these parametrizations are
universal -- a feature that can then be checked by going to more complicated
geometries.

In ref.\cite{GMP2} the above programme has already been carried out for four
quarks at the corners of a rectangle in a lattice plane. The present paper, in
which
the accuracies of the two- and four-quark energies ($V_{qq}$ and $E_i$) are
checked and also  another simple four-quark geometry is considered,
 is -- in many ways -- a follow-up to ref.\cite{GMP2}. In section 2, a two-body
potential is extracted that is more consistent than the one used in
ref.\cite{GMP2}. In section 3, the dependencies of the $E_i$ on the size of
the lattice and also on the lattice spacing are checked. The outcome is that
both the earlier $V_{qq}$ and $E_i$ of ref.\cite{GMP2} are, in general,
sufficiently accurate for the present purposes. In section 4, a short review
is given of the model proposed in refs.\cite{MGP}-\cite{GMP2}.
In section 5, the programme of ref.\cite{GMP2} is repeated for the colinear
case of four quarks on a straight line. Section 6 contains some conclusions
and future prospects.

\section{ Two-body potentials}
\setcounter{equation}{0}

\vskip 0.5 cm

The form of the heavy quark potential displayed in eq.(\ref{Vqqr}) is a
theoretical
one. It incorporates both the perturbative limit at short distances and the
linear confinement limit. However, modifications to this picture are expected
from the creation of quark-antiquark pairs. Indeed, heavy quark spectroscopy
of the $c\bar{c}$ and $b\bar{b}$ systems, which is probing mainly the
range between 0.1 and 1.0 fm, seems to favour potentials of slightly different
shape \cite{buch}. The lattice calculations to be described in this article
are done in the quenched approximation, which ignores closed quark loops.
Therefore, one expects the form of eq.(\ref{Vqqr}) to be a reasonable
approximation, if one works throughout with the
lattice data instead of data from experiment.

The lattices to be measured are generated by the standard Wilson action
in Euclidean time
for SU(2) with $\beta$=2.4 and $\beta$=2.5 using $16^3 \times 32$ and
$24^3 \times 32$ lattice points respectively to keep the physical volume
approximately constant -- for a more detailed discussion see section 3.
 A heat bath method is employed for the
equilibration of the lattice.
Further updates are combinations of three over-relaxation sweeps and one
heat bath sweep. The measurement of the appropriate correlation functions
takes place always after such a cycle of four sweeps.
In order to make error estimates, successive
measurements are collected into a series of ``blocks''. For example, later
in this section each block will consist of 240 sweeps i.e. 60 measurements
per block, and in all 12 blocks will be created.

The potential between static fundamental colour sources is obtained
from the ratio
\be
V_{qq}(r)=-\frac{1}{a} \lim_{t \rightarrow \infty} \frac{W(r,t)}{W(r,t-a)}
\ee
where $W(r,t)$ is the  Wilson loop defined as the trace of a closed loop of
link variables $U$ (encircling a rectangle of size $r \times t$ ) and
$a$ is the lattice spacing. To get accurate results
with reasonably small time separations the spatial links are replaced
iteratively by smeared or fuzzed links \cite{CM1,CM2}
\begin{eqnarray}
U^0_\mu (n) & \rightarrow & U^1_\mu (n)=A^1_n \left[ cU^0_\mu (n)+
\sum_{\stackrel{\pm \nu \neq \mu}{\nu \neq 4}} U^0_\nu (n) U^0_\mu
(n+\hat{\nu})
U^{0 \dagger}_\nu (n+\hat{\mu}) \right], \nonumber \\
U^1_\mu (n) & \rightarrow & U^2_\mu (n)=A^2_n \left[ cU^1_\mu (n)+
\sum_{\stackrel{\pm \nu \neq \mu}{\nu \neq 4}} U^1_\nu (n) U^1_\mu
(n+\hat{\nu})
U^{1 \dagger}_\nu (n+\hat{\mu}) \right], \ldots
\end{eqnarray}
i.e. each spatial link gets replaced by a multiple of itself plus a sum
of the neighbouring spatial staples. Here the $A^i_n$ are normalization factors
chosen to project the $U^i_\mu (n)$ into SU(2) and $c$ is a free parameter
adjusted together with the level of fuzzing to optimize the path overlaps
with the ground state. In the current calculations $c=4$ has been used with
the fuzzing levels 12, 16 and 20.

In the calculation of V$_{qq}$(r) the different paths ($P_i$) between quarks
 obtained by different levels of fuzzing form a variational
basis for extracting the ground state and the first excited state
\cite{CM1,CM2}. The correlation between paths $P_i$ and $P_j$ parallel
transported by $N$ steps in the time direction is
\be
W^N_{ij} = <P_i| \hat{T}^N | P_j>,
\ee
where $\hat{T} = \exp {(-a\hat{H})}$ is the transfer matrix for one time
step with the Hamiltonian $\hat{H}$, and the paths $P_{i,j}$ are
constructed as products of fuzzed links. A trial wave function
$\psi = \sum_{i} a_i |P_i>$ leads to the eigenvalue equation
\be
W^N_{ij} a^N_j = \lambda^{(N)} W^{N-1}_{ij} a^N_j,
\ee
where $\lambda^{(N)} \rightarrow \exp{(-aV_{qq})}$ as $N \rightarrow \infty $
and which, for a single path, reduces to the form given in eq. (2.1).
Besides giving the possibility to extract also higher eigenvalues
in addition to the ground state, the method yields accurate information
about V already for small values of $N$ ($N < 5$) \cite{Hunt}.
It should be added that the higher eigenstates all have the same symmetry
as the ground state. In refs.\cite{CM1,CM2} they are refered to
as $A'_{1g}$ and are not those having $E_u$ or $A_{1u}$ symmetry.
Further discussion of the different technical details can be found in
refs. \cite{CM1,CM2,Hunt}.

The results of the 720 measurements  for $\beta=2.4$ (in 12 blocks of
240 sweeps each) are displayed in Table 1
and are denoted by $av_{MC}$ with {\bf r}=($x,y,0$). The off-axis
results ($y \neq 0$) correspond to L-shaped separations of the static
sources. When a fit to these data with a potential of the form given
in eq.(\ref{Vqqr}) is attempted, it must be remembered that the lattice has
a cubic rotational symmetry and so full rotational invariance is not
guaranteed. It is known that at larger $r$-values this symmetry is
restored to a good approximation at the present values of $\beta$.
At smaller values of $r$, the lattice effects can largely be taken into
account by replacing the standard Coulomb form by the lattice Coulomb
potential \cite{LR}
\begin{eqnarray}
\label{coul}
\left(\frac{e}{r}\right)_L = \frac{e \pi}{L'^3}
\sum_{\bf q} \frac{\cos{({\bf q} \cdot {\bf r})}}{\sum_{i=1,3}
\sin^2{(aq_i/2)}} \nonumber \\
aq_i=0,\frac{2\pi}{L'},\ldots,\frac{2\pi(L'-1)}{L'} \;\; {\rm and} \; \;
{\bf q} \neq 0,
\end{eqnarray}
where $L'=2L$ and $L$ is the number of spacial sites.
 Eq. (2.6) gives the potential only upto an additive constant
and, therefore, it is normalized by fixing it to the usual Coulomb
potential $e/r$ at some large value of $r$ such that the lattice artefacts
are expected to be negligible. Here the point {\bf r} = (6a,0,0) is chosen.
The potential to be used in the fits has the form
\be
\label{coul1}
v_L(r_{ij})=-\left( \frac{e}{r_{ij}} \right)_L + b_S r_{ij} +v_0,
\ee
where $r_{ij}$ is the distance between the lattice sites $i$ and $j$.
For the first $\chi^2$ fit, the ground state energies at $x/a = 2, \ldots ,6$
and $y=0$ are used as input. For the minimization, the CERN library
programme MINUIT \cite{JR} is employed. The resulting values for the
potential parameters for $\beta=2.4$ are
\begin{displaymath}
e=0.255; \; \; b_S a^2_{2.4}=0.0691; \; \; v_0 a_{2.4}=0.560
\end{displaymath}
with $\chi^2$=0.33 per degree of freedom. The corresponding lattice
potential ($v_L$) values
for $\beta=2.4$ are given in Table 1. There it is seen that $v_L$ values
are typically within 0.3 \% from the measured values $v_{MC}$ except for
$x/a=1$, where the lattice artefacts play an important role. The second
fit combines data from the $\beta=2.4$ and $\beta=2.5$ measurements. For
$\beta=2.5$ a much smaller sample of 10 blocks of 12 sweeps was collected.
The fit involves 5 parameters, because the $e$-value is common, and only
on-axis points with $x/a=2, \ldots ,6$ are taken as input. The result
of the fit is ($\chi^2$=0.86 per degree of freedom)
\begin{displaymath}
e=0.249; \; \; b_S a^2_{2.4}=0.0699; \; \; v_0 a_{2.4}=0.555;
\; \; b_S a^2_{2.5}=0.0338; \; \; v_0 a_{2.5}=0.549
\end{displaymath}
and Table 1 lists the corresponding potential values (for $\beta$=2.4).
In this case the
typical precision is about 0.5 \% for both sets of data. The $b_S$ values
can be used to extract the ratio of the lattice spacings with the result
$\rho=a_{2.4}/a_{2.5}=1.437(17)$, where the error is an underestimate, because
correlations between different $r$-values are ignored. Furthermore, if the
reasoning given in the Introduction allows for the identification of $b_S$ with
the experimental value 0.194 GeV$^2$ \cite{CM1} for the string tension,
the lattice spacing takes on the value $a_{2.4}\approx 0.12$ fm. The inclusion
of the off-axis points ($y=0, \ldots,3$) makes the $\chi^2$-value worse
(1.29 per degree of freedom), but the parameter values remain the same within
errors, e.g.
$a_{2.4}/a_{2.5}=1.432(9)$. The fact that an acceptable fit is found with a
common value of $e$ implies that the deduced continuum  potential $v(r)$ will
be the same at both $\beta$-values. An extension
to SU(3) of this type of an analysis can be found in ref. \cite{CM2}.

Later in sections 4 and 5, the interquark potential $V_{qq}$
is needed as input for a model to describe the energies of four-quark
configurations. The optimal situation would be to use only $v_{MC}$.
However, as seen from table 1, these are not calculated for
the largest values of $x,y$. Therefore, as a compromise, $v_{MC}$
is used for $x/a\leq 3$ with $y/a\leq 3$, and for larger values of $x$ and
$y$ the  fit $v_L(\beta=2.4)$  is taken. Other alternatives, such as
the use of $v_L$ for all values of $x$ and $y$, or the combined fit
to the $\beta=2.4$ and 2.5 Monte Carlo results, would lead to results that
do not differ in any significant way from those given by the above compromise.

\section{ The energies of four quarks in rectangular geometries}
\setcounter{equation}{0}

\vskip 0.5 cm

In ref.\cite{GMP2} the energies ($E_i$) of four quarks in the coplanar
rectangular
configurations shown in fig.1 were calculated with a $16^3\times 32$
lattice and $\beta=2.4$. These values of $E_i$ are {\em exact} energies
for the four-quark system on this particular lattice. However,  the results of
physical  interest are those ($E^c_i$) for the continuum system, which should
correspond to the case of an infinitely large lattice with an infinitely
small lattice spacing $a$. Any difference between the calculated $E_i$
and the desired $E^c_i$ is a so-called lattice artefact.
To check on such possible lattice artefacts in
the results of ref.\cite{GMP2}, two points should be studied more carefully.
\begin{itemize}
\begin{enumerate}
\item To what extent
are the results dependent on the size of the lattice

being $16^3\times 32$?
\item  To what extent
are the results dependent on the lattice spacing $a$, which can be varied
by changing the coupling through changing $\beta$?
\end{enumerate}
\end{itemize}
As mentioned in the introduction, the $E_i$ are the differences  between
the total four-quark energies $[V_i(4q)]$ and that of two separated
two-quark systems $[V_{qq}]$. In the present case of rectangles with sides $r$
and
$d$ -- with $r\geq d$ -- , this reduces to
\be
\label{v4q}
E_i=V_i(4q)-2V_{qq}(d).
\ee

\subsection{ Check on finite size of lattice}
\vskip 0.5cm

At first sight it may seem that finite size effects should appear
when the distance between the quarks $(r)$ is comparable to (or
greater than) $L/2$, since in this case there is also the possibility of
 joining the two quarks with a flux path of length $L-r$, which
{\em encircles} the spatial periodic boundary. However,
 for the 2-body potential the finite size effects are known to be
relatively small -- indeed it is feasible to measure $V_{qq}(r)$
for values of $r$ greater than $ L/2$. This arises, since there is
no mixing between the direct path
with flux of length $r$ and the above one of length $L-r$. This lack of
mixing comes from the
Z(N) invariance of the lattice Wilson action in the quenched
approximation -- see appendix A and ref.\cite{Hunt} for more details.

 Now for the 4-body case in SU(2), this Z(2) invariance does not forbid direct
mixing, because two flux paths then encircle the spatial boundary.
 Thus, a priori, there may be finite size effects at
$r \approx L/2$. To estimate the size of such effects is less
straightforward but the following calculations and arguments indicate that
they are relatively small.

In order to check the dependence of the earlier results on the size of the
lattice, the latter is increased from $16^3\times 32$ to $24^3\times 32$.
{}From a numerical point of view this is a considerable change, because it
involves the storage of over $28\times 10^6$
bytes of data compared with only $8\times 10^6$
earlier i.e. approximately an increase by  the expected  factor of
$(24/16)^3$.
 Since the computer programme is constructed in such a way
that this data is read-from and written-to files many times, the numerical
efficiency drops considerably with one measurement taking about 1700 seconds
compared with about 500 seconds earlier.
In view of this, only 20 blocks of 32 sweeps  are carried out for the square
configurations i.e. $r=d$. This results in 160 measurements per
configuration since the lattice is updated by three
over-relaxation sweeps plus a heat bath sweep, which is then followed by a
measurement of the appropriate correlation functions i.e. the ratio of
up-date sweeps to measurements is 4:1.
 This number of measurements turns out to be sufficient, since the
main aim is to see whether or not the {\em average} of each energy
changes -- less attention being paid to minimizing the errors associated with
each energy.  The finite size effects are checked only for {\em squares} with
sides upto $r=7a$, since the neighbouring rectangles have binding energies
that are at least an order of magnitude smaller and -- relatively -- have
much larger error bars.
The results are given in table 2 and fig.2.

In table 2 the columns $E^{16}_i(1600)$ show the ground $(i=0)$ and
excited $(i=1)$ state energies for the 1600 measurements on the
$16^4\times 32$ lattice of ref.\cite{GMP2}, whereas those headed
$E^{24}_i(160)$ are for the 160 measurements on the present
$24^3\times 32$ lattice.
The errors $(\epsilon )$ in the two sets of
measurements are seen to be, in general, consistent with each other, since
they should be approximately related by the expression
\be
\epsilon ^{24}(N_{24})\approx\epsilon ^{16}(N_{16})
\sqrt{\frac{N_{16}16^3}{N_{24}24^3}}\approx2\epsilon ^{16}(N_{16}) .
\ee
The square root  factor is simply the ratio of the total
number of points used in each case i.e. the product of the number of
measurements times the number of points per measurement. This expression
indicates that the errors in both cases are comparable -- with the tendency
for $\epsilon^{24}$ to be somewhat larger than $\epsilon^{16}$.
This feature is qualitatively observed in table 2.

It is now seen that both the ground state energies $(E_0)$ and that of the
first excited state $(E_1)$ are in agreement with the earlier results -- the
only possible exception being $E_0(r=d=7a)$. However, there are two possible
reasons for the latter apparent difference.  Firstly, there is a purely
numerical reason. When the actual contributions from each lattice are
studied, it is found that -- for the largest two squares -- the error
distributions  are non-gaussian, so that it is not
meaningful to ascribe any significance to the discrepancies between the
results for the $16^3$ and $24^3$ lattices.
 Secondly, a difference may not be unexpected,
since there are effectively two basic energies involved with a square of
size $r=d$.
\begin{itemize}
\begin{enumerate}
\item The unperturbed energy of state A or B i.e. $E_A=2V_{qq}(r)$ -- see
fig.1.
\item The energy of two flux tubes that go to the edge of the lattice
and then, because of the cyclic boundary conditions imposed, reappear on the
opposite edge. This state (F) has an unperturbed energy of

 $E_{F}=2V_{qq}(L-r)$.
\end{enumerate}
\end{itemize}
These possibilities are shown in fig.3.

For large $r$ these energies are dominated by the linear term
$b_S$ in eq.(\ref{Vqqr}). Therefore, the $16^3\times 32$ lattice (i.e. $L=16a$)
gives for $r=7a$ the energies

\noindent $E_A=2V_{qq}(7a) \ \ {\rm and} \ \ E_F=2V_{qq}(9a)$
with the difference $E_F-E_A\approx 4b_Sa \approx 0.3/a $.
 On the other hand, for the $24^3\times 32$ lattice
$ E_F=2V_{qq}(17a)$ with the difference
$E_F-E_A\approx 20b_Sa \approx 1.5/a$.
Having made this observation, it is still not easy to calculate accurately
the effect the F-states have had on the energies $E_0$ and $E_1$.

In appendix B, this discussion is pursued further in the framework of the
model to be introduced in section 4. There arguments are given that suggest
 the finite size correction is, indeed, unimportant for
$E_0(r=d=7a)$ on the $16^3 \times 32$ lattice.

\vspace{0.3cm}

\subsection{ Check on scaling}
\vskip 0.2cm
So far in this paper, all the four-quark measurements have been carried out
with
$\beta=2.4$. At this $\beta$ value the string tension
$b_Sa_{2.4}^2=0.0699$  in
eq.(\ref{Vqqr}), which corresponds to a lattice spacing of $a\approx 0.12$ fm.
 For the present results to be meaningful, since any self-energies cancel,
the values of the four-quark energies ($E_i$) should be
independent of $a$. In other words, for two different lattice spacings
$a_1$ and $a_2$ with $\rho=a_1/a_2$, the energies should satisfy
\be
\label{scal}
E_i(a_1,r)= E_i(a_2,r).
\ee
In order to isolate this scaling effect from a finite size effect,
it is convenient to use lattices which have approximately the  same
{\em physical} size. Since for $\beta=2.4$ the lattice spacing is 0.12 fm,
the $16^3$ lattice corresponds to $\approx 7 $ fm$^3$. Therefore, the $24^3$
lattice requires $a\approx 0.082$ fm, which corresponds to $\beta\approx 2.5$.
In practice, $\beta=2.5$ is used to give $\rho=1.437(17)$ -- see section 2 and
ref.\cite{CM1}.
As in the case of the finite size check, the use of a $24^3\times 32$ lattice
again prohibits numerically an exhaustive study. Therefore, only
11 blocks of 240 sweeps i.e. 660 measurements are made. The results are shown
in table 3 and fig.4.

Here it is seen that the hoped-for agreement expressed by
eq.(\ref{scal}) is only satisfied for the largest squares to within about
$20\%$. However, later in section 4, it will be seen that  these differences
between the results  for the  largest squares with the two values of $\beta$
 are very much smaller than the differences between these values
of $E_i$ and the predictions ($E'_i$) of a conventional model based on
pair-wise
interquark potentials i.e. $|E'_i-E_i|\gg |E_i(\beta =2.5)-E_i(\beta =2.4)|$.
For example, later in table 4 at $r=5a$ it will be shown that $aE'_0=-0.104$
and $aE'_1=0.312$, making $|E'_0-E_0|\approx 0.07/a$
and $|E'_1-E_1|\approx 0.21/a$  compared with
$|E_0(2.5)-E_0(2.4)|\approx 0.015/a(2.4)$
 and $|E_1(2.5)-E_1(2.4)|\approx 0.030/a(2.4)$.

Therefore, the conclusion to be drawn from this subsection is that, for the
present purposes,  scaling has been achieved with the value of
$\beta=2.4$. However, as discussed in ref.\cite{mich}, asymptotic
scaling -- which relates the results from differing $\beta$'s by
perturbative arguments -- should not be expected at these values of $\beta$.

\subsection{ Autocorrelation}
\vskip 0.5cm

Since the above type of simulation involves the generation of  successive
updated lattice configurations, one source of uncertainty in the error
estimation is the degree to
which these configurations are statistically
independent of each other. In the present work, the results are collected
into $N_B$ blocks each containing the average $E_i (i=1,..,N_B)$
 of $N_M$ measurements. From these numbers, the
quantities of interest are the  overall energy average, $\bar{E}$, and its
error $\sigma (\bar{E})$, which can be expressed as
\be \bar{E}=\frac{1}{N_B}\sum_{i=1}^{N_B} E_i \ \ {\rm and } \ \
\sigma (\bar{E})=\sigma (E)\sqrt{\frac{1}{N_B}f(Auto)},
\ee
where
\be
\sigma (E)=\sqrt{\frac{1}{N_B}\sum_{i=1}^{N_B}(E_i-\bar{E})^2}
\ee
is the standard deviation of the $E_i$'s and $f(Auto)$ is calculated as follows
 -- see ref.\cite{Priest}.
The normalised autocorrelation function between $E_i$ and $E_{i+k}$ is
defined as
\be
\rho(k)=\frac{C(k)}{C(0)}=
\frac{\frac{1}{N_B-k}\left[ \sum_{i=1}^{N_B-k}(E_i-\bar{E})
(E_{i+k}-\bar{E})\right]}
{\frac{1}{N_B}\left[\sum_{i=1}^{N_B} (E_i-\bar{E})^2\right]},
\ee
giving
\be
\label{f(A)}
f(Auto)=1+2\sum_{k=1}^M(1-\frac{k}{N_B}) \rho(k).
\ee
As discussed in ref.\cite{Mad}, $f(Auto)$ has the significance
of a decay time $\tau=\frac{1}{2}f(Auto)$.
The value of $M$ is then chosen to retain most of the ``signal'' but discard
most of the ``noise''. This is carried out by imposing the condition that $M$
need only be large enough to ensure it is several ($n\approx$ 3 or 4)
 times greater
than $\tau$ i.e. $M$ should be the smallest integer to
satisfy the equation $M\geq n\tau=n\frac{1}{2}f(Auto,M)$.
It is found that $M\approx 2$ in this particular problem.
The ratio $N_B/f(Auto)$ can then be interpreted as the equivalent number of
independent blocks.
In the present study, it is expected that only the correlation  between nearest
neighbours could possibly be of importance i.e. $E_i=cE_{i-1} + \epsilon_i$,
where $\epsilon_i$ is not correlated with other $E_i$'s
 and $|c|$ is assumed to be less than unity.
 In this idealised case, for large $M$ and $N_B$, this gives
$\rho(k)=c^k$ and $f(Auto)\approx 1+2(c+c^2+c^3+....)=(1+c)/(1-c)
=(1+\rho(1))/(1-\rho(1))$.

As typical illustrations of these estimates, there is shown in fig.5
the values of $\rho(k)$ for different values of $\beta$ and
different lattices. In all cases $\rho(1)$ is seen to be small
i.e. $|\rho(1)|\leq 0.2$. Also, since the limitation of $M$  in the sum
in eq.(\ref{f(A)}) is 1 or 2 to eliminate the ``noise'', it follows that
$\sqrt{f(Auto)}\approx 1.0\pm 0.2$.

The conclusion to be drawn from this subsection is that the error,
$\sigma(\bar{E})$, on the average value of the $E_i$ is adequately given by
simply
$\sigma (E)\sqrt{\frac{1}{N_B}}$ i.e. with the autocorrelation effect
expressed by $f(Auto)$ being negligible.

\section{  Review of model}
\setcounter{equation}{0}

\vskip 0.2 cm
So far two of the three original requirements listed in the introduction
have been calculated -- namely -- accurate and reliable values of
four-quark binding energies ($E_i$) and also the corresponding interquark
potential $V_{qq}$. It now remains to set up a
physically motivated and convenient model or prescription
for calculating the energies $(E'_i)$ of a four-quark
system in terms of the $V_{qq}$ -- the hope being that the $E'_i$ are a good
approximation to the exact $E_i$.

As discussed in refs.\cite{MGP}-\cite{GMP2}, the first model that comes to
mind is one in which configurations A and B in fig.1 are used as a basis for a
two-by-two matrix constructed directly from the potentials $V_{qq}$. In the
static approximation, for which the kinetic energy is zero, eigenvalues
$\lambda_i$ can then be obtained by diagonalizing
\be
\label{Ham}
\left[{\bf V}-\lambda_i {\bf N}\right]\Psi_i=0,
\ee
with
\be
\label{NV}
{\bf N}=\left(\begin{array}{ll}
1&1/2\\
1/2&1\end{array}\right)\ \ {\rm and}\ \ {\bf V}=\left(\begin{array}{cc}
v_{13}+v_{24} & V_{AB}\\
V_{BA}&v_{14}+v_{23}\end{array}\right),
\ee
where $v_{ij}=V_{qq}(r_{ij})$ and
\be
\label{AVB}
<A|V|B>=V_{AB}=V_{BA}=
\frac{1}{2}\left(v_{13} +v_{24} +v_{14}+v_{23} - v_{12}-v_{34} \right).
\ee
Here the colour factors are first determined using the form of $v_{ij}$
that is pure one-gluon-exchange. Only after this are the radial forms of
$v_{ij}$ taken from table 1 or eq.(\ref{coul1}) -- see ref.\cite{GMP2}.
The matrix {\bf N} arises since the configurations A and B, when expressed
purely
in terms of quark degrees of freedom, are not orthogonal but have the
overlap $<A|B>=1/2$. In general for SU(N), this overlap is 1/N.
{}From the eigenvalues $\lambda_i$ are now subtracted the internal energy
of the meson-like state with the lowest energy [i.e. $v_{13}(d)+v_{24}(d)$ for
$d\leq r$]. This gives the quantities
\be
E'_i(d,r)=\lambda_i-[v_{13}(d)+v_{24}(d)]
\ee
 that can be interpreted
 as the binding energy of two mesons of length $d$ a distance $r$ apart.
Since a very similar subtraction is also made in the lattice calculation
of section 3,
it is meaningful to compare $E'_i(d,r)$ with the $E_i(d,r)$ in tables 2 and
3. This is shown in table 4, where
it is seen that the $|E'_i|$ are always greater than the $|E_i|$.
The difference increases as $r$ or $d$ increase until, for the largest
squares, considered $|E'_i| \approx 10|E_i|$.

To overcome the above differences between $E_i$ and $E'_i$, in
refs.\cite{MGP}-\cite{GMP2}
a modified version of the above model was proposed, in which the four-quark
energies $E_i$ could be understood in terms of the interquark potential
$V_{qq}$ and a quantity $f$ -- a phenomenological function of the spacial
coordinates of all four quarks. This function can be interpreted as an overlap
factor of the gluon fields mediating the interaction between the quarks.
Using again the configurations A and B in fig.1 as a basis, the problem now
reduces to diagonalising
\be
\label{Hamf}
\left[{\bf V}(f)-\lambda_i(f) {\bf N}(f)\right]\Psi_i=0,
\ee
with
\be
\label{NVf}
{\bf N}(f)=\left(\begin{array}{ll}
1&f/2\\
f/2&1\end{array}\right)\ \ {\rm and}\ \ {\bf V}(f)=\left(\begin{array}{cc}
v_{13}+v_{24} & fV_{AB}\\
fV_{BA}&v_{14}+v_{23}\end{array}\right).
\ee
{}From the eigenvalues $\lambda_i(f)$ are again subtracted the internal energy
of the meson-like state with the lowest energy [i.e. $v_{13}(d)+v_{24}(d)$ for
$d\leq r$]. This gives the quantities
\be
E'_i(f;d,r)=\lambda_i(f)-[v_{13}(d)+v_{24}(d)].
\ee
To ensure the equality $E'_i(f)=E_i$, the function $f$ is varied. In
practice, it is found that -- for each configuration $(d,r)$ -- a single
value of $f$, to be denoted by $\bar{f}$,
 is able to give an optimal fit to both $E'_0(\bar{f})=E_0$
and  $E'_1(\bar{f})=E_1$, indicating that the parametrization suggested
in eqs.(\ref{Hamf},\ref{NVf}) contains the most important features of
the more precise lattice calculation. The results are shown in table 4.

One important consequence of the above fit of $E'_i(\bar{f})$ to $E_i$ is that,
for
the larger squares, $\bar{f}$ is certainly {\em not} consistent with unity.
This means that
conventional ways \cite{cluster} of treating the interaction between quark
clusters purely in terms of the pair-wise interquark potential of
eq.(\ref{Vqqr}) are dubious.  However, as the size of the squares becomes
smaller, the value of $\bar{f}$ tends towards unity. For the smallest square
considered in this work i.e. $r=d=0.082$ fm -- the (1,1) case with $\beta=2.5$
in table
3 --
$E'_0(f=1)=-0.0626/a(2.5)$, $E'_1(f=1)=0.1878/a(2.5)$ and $\bar{f}=0.92$. This
is in
line with the expectation that the weak coupling limit
(i.e. one-gluon-exchange corresponding to $f=1$) emerges as $r$ and $d$ tend
to zero.

In table 4, for non-squares, the values of $\bar{f}$ are dictated mainly
by the energies of the excited states ($E_1$). The reason for this is clear
from the following argument. Use the notation of fig.1 and also

$v_1=v_{13}(d)=v_{24}(d), \ v_2=v_{14}(r)=v_{23}(r), \ v_d=v_{12}=v_{34}
, \  \Delta=2(v_2-v_1) \\ {\rm and} \ \bar{\Delta}=2(v_2-v_d).$
It is now seen that $\bar{\Delta}/\Delta\rightarrow 0$ as the rectangles become
more and more elongated. In this case, by expanding in powers of
$\bar{\Delta}/\Delta$,
\be
\label{elong}
E'_0\approx -\frac{\bar{f}^2\bar{\Delta}^2}
{4\Delta}  \ \ {\rm and} \ \
E'_1\approx\frac{\Delta-\frac{\bar{f}^2\bar{\Delta}}{2}}
{1-\frac{\bar{f}^2}{4}}.
\ee
In other words, to lowest order, when the coupling ($\bar{\Delta}$) vanishes,
$E'_0\approx 0$ and $E'_1\approx \Delta/ (1-\frac{\bar{f}^2}{4})$
i.e. $E'_1(\bar{f})=E_1$ gives a direct
measurement of $\bar{f}$. For example, even with the $d=a, \  r=2a$ case, from
table 1, it is seen that
$a\Delta=0.3766 $ and $a\bar{\Delta}=-0.0806$. The approximate expressions
in eqs.(\ref{elong}) then give -- for the optimal $\bar{f}=0.869$ from
table 4 -- the values
$aE'_0\approx-0.0030$ and $aE'_1\approx 0.502$ in good agreement with the Monte
Carlo results. This suggests that for non-squares it may be better to
concentrate on extracting the energy of the first excited state, since
this gives a much more reliable value of $f$ than $E_0$.
However, this is not without its problems. Firstly, in the present
lattice calculation the highest eigenvalue extracted $(E_1)$ is
influenced more than $E_0$ by the presence of even higher eigenstates.
Secondly, in the model proposed in eqs.(\ref{Hamf},\ref{NVf}) only two
configurations A and B are explicitly introduced to explain $E_0$ and $E_1$.
However, these two configurations and the factor $f$ may not be simulating
the effect of other configurations ( e.g. glue-ball exchange) in the same way
as these effects appear in the Monte Carlo energies
$E_0$ and $E_1$. Again, the higher eigenvalue
$E_1$ would be expected to be the less reliable.

The existence of a function $f$ that depends on the positions of all four
quarks is of interest but not particularly useful, unless it can be
parametrized in some convenient manner. In refs.\cite{MGP}-\cite{GMP2}
two such parametrizations were tested.
{}From refs.\cite{Morimatsu,Morimatsu2}
-- motivated by strong coupling arguments -- the phenomenological
form was taken to be
\be
\label{f1}
f_1=\exp[-\alpha b_S S],
\ee
where $b_Sa_{2.4}^2=0.0699$ is the string tension dictated by the interquark
potential of eq.(\ref{Vqqr}) and $S$ is the minimal area of the surface
bounded by the straight lines connecting the quarks and antiquarks.
The other form -- the one proposed in \cite{Masud} -- was
\be
\label{f2}
f_2=\exp[-\frac{kb_S}{6}\sum\limits_{i<j}r^2_{ij}]
\ee
i.e. the cut-down is governed by the average of the six links
present in a $q^2\bar{q}^2$ system.
In eqs.(\ref{f1},\ref{f2}) the $\alpha$ and $k$ are at present free
parameters to be determined by fitting $f_1$ and $f_2$ to the $\bar{f}$ in
table 4. It is then of interest to study whether either $\alpha$ or $k$ is
constant as the values of $r$ and $d$ vary.

Both of these parametrizations of $f$ accommodate the two
extreme models of:

1) Weak coupling, which has $f=1$ when all $r_{ij}=0$ .

2) Strong coupling, which has $f=0$ when any $r_{ij}\rightarrow \infty$.

\noindent For squares, these two parametrizations are equivalent with
$k=3\alpha/4$ -- so that only the few cases with $r\not =d$
( not included in the finite size and scaling tests of tables 2 and 3) could
possibly distinguish between $f_1$ and $f_2$. From ref.\cite{GMP2}, the
conclusion was that $\bar{k}=3\bar{\alpha}/4=0.5\pm 0.1$ achieved the best
overall fit to $\bar{f}$ -- with the error bars on the $E_i$ for the non-square
configurations being too large to separate $f_1$ from $f_2$.
However, with  the use of the more consistent set of $V_{qq}$ from table 1 and
also the realisation that fitting $E'_1(\bar{f})$ to $E_1$ is possibly more
informative for extracting a value of $\bar{f}$, the new results for $\bar{k}$
and
$\bar{\alpha}$
are shown in table 4. Here it is now seen that -- for squares --  $\bar{k}$ (
and
therefore $\bar{\alpha}$) shows a distinct tendency to decrease as the
squares become larger. Furthermore, for rectangles the values of
$\bar{\alpha}$ appear to decrease less for a fixed $d$ with increasing $r$.
This
suggests that the $f_1$ parametrization -- although not optimal -- is
superior to that of $f_2$.

Clearly, to get a better understanding of $f$, more configurations very
different from squares are necessary. Within the framework of
rectangles, geometries with $r\gg d$ would be suitable. However, as shown
 in table 4,

$E_0(d,r=d)\approx 10\times E_0(d,r=d+a)  \approx 100\times E_0(d,r=d+2a)$,

\noindent which rules out extracting $E_0$ for more elongated
geometries, since already
the results for $E_0(d,r=d+2a)$ are comparable to the error bars e.g.

$aE_0(1,1)=-0.0695(5),\ \ aE(1,2)=-0.0026(2), \ \ aE(1,3)=-0.0003(3)$.

\noindent However, as shown in eq.(\ref{elong}), in these cases it is $E_1$
that is of more importance.
Therefore, one extension to the above analysis could be to concentrate on $E_1$
 for configurations with $r\gg d$. In such cases, $f$ would be given
directly by eq.(\ref{elong}) as
$\bar{f}\approx 2\sqrt{1-\Delta/E_1}$. The restriction on $r$ would then
only be limited by the finite lattice size problem discussed earlier.
This possibility will not be considered here. Instead, in the next section
the even more extreme geometry of four colinear quarks will be studied.

\vspace{0.3cm}

\section{  The case of four colinear quarks}
\setcounter{equation}{0}

\vskip 0.2 cm
Another series of four-quark configurations -- that are very different from
rectangles and yet can be treated easily by a Monte Carlo simulation -- is
the geometry, shown in fig.6 of four quarks on a straight line.

In common with the configurations A and B in fig.1, all of the basic states
are constructed from paths that essentially involve a {\em single} spatial
dimension. For example, since they represent mesons in their ground state,
$M_{13}$ and $M_{24}$ in A can be constructed from links all in the
$y$-direction. Likewise, $M_{14}$ and $M_{23}$ need only links in the
$x$-direction. If the rectangular state
$C=M_{12}M_{34}=[q_1q_2]^0[q_3q_4]^0$ had been included, then L-shaped paths
involving both $x$- and $y$-links would have been necessary. In fact, in
table 1  such paths were considered in order to extract the
two-quark potential $V_{qq}(x,y)$ for $y\leq x$ -- quantities needed in
eq.(\ref{AVB}) when setting up the model of eqs.(\ref{Hamf},\ref{NVf}).
In ref.\cite{GMP2}, it was considered sufficient to  treat only the two
configurations
A and B in fig.1, since state C has a higher energy. However, in the linear
case the distinction between states A,B and C is less clear, since -- in the
domain where the two-quark potential of eq.(\ref{Vqqr}) is dominated
by the $b_Sr$ term -- states B and C have the {\em same} energy which
is always greater than that of state A.
 At this point,
it should be remembered that the four quarks are, in fact, fermions and
a more correct notation for states A,B and C is in terms of quarks and
{\em antiquarks}.
The
use throughout of SU(2) means that
 quarks and antiquarks belong to the same two dimensional
 representation of the colour group. However, in any discussion
 involving all three states A,B and C, it is important -- even with
SU(2) -- to keep track of phases.  It should be noted that,  in the limit when
all link matrices U are unit matrices,  the three off-diagonal overlaps of the
states A, B, and C are all $\pm 1/2$ where the signs are convention-dependent;
however,  independent of convention,  the product of all three signs is
negative.  This minus sign must be put in by hand in the $3\times 3$
calculation.

In ref.\cite{GMP2} the use of A and B as a basis in the Monte Carlo
simulation was considered from a variational point of view. In
principle, using only state A or B would lead to the same ground state
energy $E_0$, but this would require extrapolating further in Euclidean
time T. In addition, the extended variational basis of A and B enabled an
estimate to be made of the excited state energy $E_1$. In the light
of this discussion, it might be
thought that the even larger basis  of A, B and C would lead to a more rapid
convergence in T and also give the three lowest energies $E_{0,1,2}$.
However, because of the way -- in this work -- the states A, B and C are
constructed
from basic one-dimensional lattice links, for the case of colinear
configurations
these states are not linearly independent of each other. Therefore, the
inclusion of all three leads to a singular matrix in the numerical evaluation.
 This singularity would not arise in general.

\subsection{ Results of the Monte Carlo simulation and their interpretation
in terms of the model in section 4}

\vskip 0.2cm

In the notation of fig.6, the colinear geometries considered have
the symmetric form $r_{13}=r_{24}=d$ and $r_{12}=r$ with $d/a=1,2,3$ and
$r=d+a,d+2a,d+3a$. As discussed in section 3-A,
this means that for $d=3a$ and $r=5a, \ 6a$ finite lattice artifacts could be
playing a role, since in these cases flux paths of length $8a$ and $9a$ enter.

The results for 20 blocks of 320 sweeps each, i.e. 1600 measurements, on
a $16^3\times 32$ lattice with $\beta=2.4$ are shown in tables 5 and 6.

Several points can be made about these tables.

1) $E_0(r,d)$ is  significantly non-zero only for $r=d+a$ i.e. for those
configurations in which the two mesons are within one lattice spacing
at their point of closest approach.

2) $E'_i(f=1)\not= E_i$ implying that $f\not=1$. Therefore, again -- as
in the rectangle case -- the conventional two-quark potential model
corresponding to $f=1$ is not successful.

3) Because $f\not=1$ is required, this implies the parametrization
$f_1$ in eq.(\ref{f1}) cannot be correct, since for linear configurations,
the area $S$ in the definition of $f_1$ is always zero -- to give
$f_1=1$ for all linear cases.

4) Since the accuracy of $E_1$ is considerably better than that for $E_0$, the
compromise value of $\bar{f}$ ensures $E'_1(\bar{f})=E_1$ to three significant
figures. On the other hand, the $E'_0(\bar{f})$ are less precisely
determined but still remain within the error bars of the $E_0$. The reason
for this was given in eq.(\ref{elong}). However, as discussed after that
equation, the use of $E_1$ does have problems not present with $E_0$.

5) As seen from fig.7, the values of $\bar{f}$ needed to fit $E_i(\bar{f})=E_i$
are
remarkably {\em constant} with the value $\bar{f}=0.86(2)$ covering the
whole range of colinear geometries considered here. This indicates that
it will be difficult to extract from $\bar{f}$ any precise geometry
dependence of the form suggested in eq.(\ref{f2}). Indeed,
the values of $\bar{k}$ needed to satisfy
$\bar{f}_2=\exp\left(-\bar{k}b_s<r_{ij}^2>\right)$ show a clear tendency to
decrease as the four-quark system gets larger -- see table 6 and fig.7.
The use of $(r,d)$ as the abscissa is not the most illuminating, since
then the value of $<r_{ij}^2>$ -- a more appropriate variable -- is not evenly
spaced. In fact, for (4,1) the value of $6<r_{ij}^2/a^2>$  is 68, whereas for
(3,2) it is only 52. However, this does not alter the conclusion that
$\bar{k}$ is {\em not } constant.

6) Another alternative parametrization of $f$ is
$\bar{f}_3=\exp\left(-\bar{p}\sqrt{b_s}L\right)$, where $L$ is the length
of the perimeter enclosing the area $S$ of $f_1$ in eq.(\ref{f1}).
 For the present colinear
case, $L=2(d+r)$ and the corresponding values of $\bar{p}$ are also given
in table 6 and figure 7. However, as expected from the constancy of
$\bar{f}$, this does not fare any better than the previous parametrization
$\bar{f}_2$ with $\bar{k}$. Again $\bar{p}$ has a clear tendency to decrease as
the
size of the system increases. Here a more appropriate abscissa variable
 would be the perimeter $L$ when plotting figures.  In this case the plot
of $\bar{p}$ would be somewhat distorted
with, for example, (4,1) and (3,2) corresponding to the same abscissa. But
again, this does not alter the conclusion that $\bar{p}$ is {\em not} constant.

The overall conclusion to be drawn from the above is that  a
reduction factor $\bar{f}$ is needed -- but that this factor is essentially
constant
at $\bar{f}=0.86(2)$ for all of the colinear geometries considered.

So far the Monte Carlo results and their subsequent analysis in terms of
the model reviewed in section 4 have been purely in terms of the states
A+B in fig. 6. However, similar analyses can be carried out with
A+C or B+C. As expected, in the Monte Carlo simulation these lead to exactly
the same values of $E_{0,1}$ in table 5. Also the three model energies
$E'_{0,1}$$(f=1$,A+B) \ , \ $E'_{0,1}$$(f=1$,A+C) and $E'_{0,1}$$(f=1$,B+C) are
all equal. But this symmetry does not persist, when the corresponding values
of $f$ are extracted to ensure $E'_i(f_{AC})=E_i$ or $E'_i(f_{BC})=E_i.$
This is shown in table 7, where the values of $f_{AB},f_{AC}$ and $f_{BC}$
 are given to ensure $E'_0(f)=E_0$ and in the brackets [...] to
ensure  $E'_1(f)=E_1$. However, in this case the use of the B,C combination
is clearly not very sensible, since it misses the lightest state A and then
attempts to reproduce its presence.

The following points can be made about table 7.

1) Since the $E_0$'s have larger error bars than the $E_1$'s, the resultant
errors on $f_{AB}(E_0)$ and $f_{AC}(E_0)$ are also larger than those on the
$f_{AB}(E_1)$ and $f_{AC}(E_1)$. In general, within these error bars
$f_{AB}(E_0)\approx f_{AC}(E_0)$.

2)  $f_{BC}(E_0)=1.0$ to within $1\%$ -- a result readily understood,
since from eqs.(\ref{Hamf},\ref{NVf}) it follows that
$f_{BC}(E_0)\approx 1-\frac{E_0}{4(r-d)}$, when the interquark potential
$v(r)$ is simply proportional to  $r$.

3) Those values of $f$ adjusted to fit $E'_1(f)=E_1$ show much more variation
 than those fitting $E'_0(f)=E_0$. Also, in some cases, no solution exists
for any value of the overlap $f$.

4) For most geometries, in cases A+B and A+C, a compromise value of $f$ can
be found that gives reasonable fits to both
$E_0$ and $E_1$. However, for B+C, the $f$'s that ensure
$E'_1(f(E_1))=E_1$ yield values of $E'_0(f(E_1))$ that are very far from
$E_0$ e.g. in geometry (3,1), $aE'_0(f=0.58)=0.352$ compared with the Monte
Carlo
number $aE_0=-0.0002(3)$. In other words, there is no compromise value of
$f$ for case B+C.

The above asymmetry between the use of A+B, A+C or B+C, when fitting
$E'_i$ to $E_i$ is not entirely satisfactory and has to be resolved, if
something like the 2$\times$ 2 model of section 4
is to be used as the basis of a dynamical
calculation as in ref.\cite{Masud}.

\subsection{ A test case of QCD in two dimensions (1+1)}

\vspace{0.2cm}

The above analysis unavoidably involves numerical results extracted
from Monte Carlo simulations on a 4-dimensional lattice. However,
 it is possible to study colinear colour sources in a simple
approximation for which exact theoretical results are known.
This is QCD in two dimensions (QCD2) -- the one spatial direction
 allowing colinear (and only colinear) configurations. For quenched QCD2,
the spectrum is known {\em exactly} even on a lattice. This
can be summarised as the requirement that each
link is in a representation of the colour group (i.e. SU(2) here).
Therefore, the links can be in the singlet ground state (J=0), or they can be
excited
to a fundamental representation (J=1/2), an adjoint representation
(J=1),.. etc. The energy per unit length for such links is given by
$b_J=\frac{4}{3}KJ(J+1)$.
For 4 colinear quarks, the lowest state (A in fig.6)
 has energy $2Kd$, while the first excited state has, in the middle, an adjoint
link of length $(r-d)$ resulting in an energy $2Kd+8K(r-d)/3$.  This exact
result
applies at any $\beta$-value, i.e. strong coupling or weak coupling.

A comparison with the above $f$-mixing model of
eqs.(\ref{Hamf},\ref{NVf}) shows that
there is  agreement
with the exact QCD2 results  provided $f=1$ -- independent of which
combination of states is used for the analysis A+B, A+C or B+C.
 This is also true in the more general case when $d_{13}\not= d_{24}$.
If the interpretation of $f$ being a gluon-field overlap factor
is correct, then it is easy
to understand that $f$ must be unity, since with only one spatial direction
the colour flux must overlap fully.
Earlier, the main motivation for the $f$-mixing model had been from
weak coupling arguments. So this agreement
between the mixing model and QCD2 does suggest that the model
is sensible even at large $\beta$. This adds support to
the claim that it may be a useful
phenomenological tool when the extension to $f\not = 1$ is made.

\section{  Conclusion}

In this paper, by considering a $24^3\times 32$ lattice and $\beta=2.5$,
 it has been shown that the use of a $16^3\times 32$
lattice with $\beta=2.4$ does not suffer from significant lattice
size or scaling effects. Therefore, the results for the smaller lattice and
$\beta$, quoted in tables 4 and 5, are expected to be a true reflection
of the four-quark energies $E_i$ in the {\em continuum} limit.

Compared with refs.\cite{GMP,GMP2} the case of four colinear quarks is in
addition considered.
Analysing the results using states A+B in fig.6, it is found that the
gluon field overlap factor $f$
is essentially constant at 0.86(2) for all of the colinear geometries
treated.

At this stage, it is not clear which parametrization of $f$ is optimal.
However, if $f$ is only needed to $\approx 15\%$ accuracy, then the
area dependent form of $f$ in eq.(\ref{f1}) appears to be superior.
Work is now in progress on other more elaborate four-quark
geometries and this, hopefully, will shed more light on a more
optimal form for $f$.

So far the Monte Carlo simulation results ($E_i$) have been interpreted by
 the $2\times 2$ model of section 4. However, as discussed at the end of
section 5, there are features of this model that can be considered as
unsatisfactory. In particular, the present formulation of the model does not
retain the symmetry between the states A,B and C that existed between the
corresponding states in the Monte Carlo simulation. In view of this, it is
 possibly too ambitious to present the $2\times 2$ analysis of section 4
as a fully justified model. It is probably more correct to think of this
analysis as simply a convenient and physically motivated
{\em prescription} for reconstructing the Monte Carlo results.

The authors wish to thank C.Montonen for useful discussions, C. Alexandrou and
M. Locher for useful comments on the manuscript and, in addition, one of the
authors (MES) thanks the Magnus Ehrnrooth Foundation for a grant.
The authors also wish to acknowledge that these  calculations were performed on
the CRAY X-MP machines at both Helsinki and RAL (UK) - computer time on
the latter being financed by a grant from SERC.

\vskip 1.0 cm

\appendix

\chapter{The effect of Z(N) invariance}
\renewcommand{\theequation}{A.\arabic{equation}}

The simplest way to describe a gauge-invariant interaction for gluon-fields
is with the Wilson action, which can be expressed in terms of a product of
four link variables $U$ as
\be
U_P=U_\nu (n) U_\mu (n+\hat{\nu})U^{ \dagger}_\nu (n+\hat{\mu})
U^{ \dagger}_\mu (n)
\ee
around a plaquette $P$, whose corners have the coordinates
$n,n+\hat{\nu},n+\hat{\nu}+\hat{\mu},n+\hat{\mu}.$
A possible transformation on $U_P$
is to multiply the two links in the $\nu$-direction
by the element $(z)$ of some group. If this is to leave $U_P$ invariant, then
$z$ must commute with the $U$'s and have the property $zz^{\dagger}=1$ i.e

\[U_P\rightarrow zU_\nu (n) U_\mu (n+\hat{\nu})(zU)^{ \dagger}_\nu
(n+\hat{\mu})
U^{ \dagger}_\mu (n)\]
\be
 =zz^{\dagger}U_\nu (n) U_\mu (n+\hat{\nu})U^{ \dagger}_\nu (n+\hat{\mu})
U^{ \dagger}_\mu (n)  =zz^{\dagger}U_P =U_P
\ee
Since the $U$'s are elements of SU(N) and the $z$'s commute with
all of these elements, then the $z$'s must be the elements of the centre
of SU(N) -- namely the group Z(N), whose elements are the Nth roots of unity.
The general form of the elements of Z(N) are
$z_q=\exp(i2\pi q/N)$ with $q=0,1,....,N-1$.

The above argument involves only a single plaquette. However, the links
 $U_\nu (n) $ and $U_\nu (n+\hat{\mu})$ are also parts of
neighbouring plaquettes. Therefore, to
ensure the total action is invariant, all links in a 3-D plane defined
by $n$ and $\hat{\mu}$ must be multiplied by the same element $z_q$.

Figure 8  shows the (x,T)-plane and the two possible flux paths connecting
quarks at $x_1$ and $x_2$ -- a direct path ($A$) of length $r$ and
a second $(FA)$ of length $L-r$ encircling the spatial boundary.
Finite size effects arise when $A$ and $FA$ have a non-zero correlation
i.e. $<FA | A>\not =0.$
Figure 8 also shows the plane in which all the $U$ links have been
transformed into $zU$. This is seen to intersect {\em only} state $A$
i.e.
\be
<FA| A> \rightarrow <FA| zA>=z<FA|A>.
\ee
Since such correlations must have the same invariance under Z(N)
as the basic interaction, this shows that $<FA| A>$ must {\em vanish}
i.e. there is no finite size effect on state $A$ due to the presence
of state $FA$. However, if the flux path in state $FA$ encircles the spatial
periodic boundary N--1 times then
\be
<FA|A>\rightarrow <(z^{\dagger})^{N-1}FA|zA>=z^N<FA|A>=<FA|A>.
\ee
Therefore, for SU(2) finite size effects are not expected to arise
until $r$ approaches $L$ -- and not $L/2$ as might have been naively thought.
 Also for SU(N), with N$>$2, the finite size
effects are expected to be smaller, since they require states $FA$
that have encircled the boundary more than once.

The above ideas are easily extended to four-quark geometries. In the present
case of SU(2), the plane in which all $U$ links have been transformed
into $zU$ is now intersected by {\em both} direct flux paths -- see
fig.3. Therefore,
\be
<FA|A>\rightarrow <FA|z^2A>=<FA|A>
\ee
i.e. finite size effects are expected already for $r\geq L/2$.

\vskip 1.0cm

\chapter{A comment on the magnitude of finite lattice size effects}
\renewcommand{\theequation}{B.\arabic{equation}}

In section 3A, the question was raised concerning the magnitude of finite
lattice size effects on the energy of the square configuration with
$r=d=7a$ on the $16^3 \times 32$ lattice.
A possible indication can be obtained by comparing this
situation with that occuring for smaller squares. As was seen in table 4
of section 4, the ground state binding energies of square configurations
(i.e. $r=d$) are approximately --0.05, whereas that of neighbouring
rectangles is much smaller e.g.

\noindent $aE_0(d=2a,r=3a)\approx $--0.006 and
$aE_0(d=2a,r=4a) \approx$ --0.001 compared with $aE_0(r=d=2a)\approx $--0.06.
This means that, in the notation of fig.1, the state B has little effect on
A, when $r \not=d$.
In the case of the large squares of interest for the finite size problem
 and being guided by the model introduced in section 4,
the analogous situation can be considered as basically four states
A, B, FA and FB interacting with each other. First consider each of these
states
 interacting in {\em pairs}.

1) A$(d,r)$+B$(d,r)$ gives a large binding for the square case $r=d=7a$.

2) A$(d,d)$+FB$(d,d)$ (or B+FA) will interact to give a binding that is the
  {\em same} as
A$(d,d)$+B$(L-d,d)$. However, since $L-d\not= d$, this binding should be
small. In fact, if the result for small values of $r,d$ is also true here, then
it is to
be expected that the binding due to A(7,7)+FB(7,7) could be upto {\em two
orders of magnitude smaller} than that due to A(7,7)+B(7,7). Similarly,
B(7,7)+FA(7,7)
should contribute little to the overall binding of the four states A, B, FA,
FB.

3) It remains to discuss the two pairs A$(d,d)$+FA$(d,d)$ and
B$(d,d)$+FB$(d,d)$. These could produce even less binding than A+FB or B+FA
for the following reasons. As discussed in ref.\cite{GMP2} and also in
section 4, for smaller values of $r$ and $d$ the non-square configurations
have less binding than the neighbouring square cases because the mixing
interaction $V_{AB}$ is much less than the difference in unperturbed energies
$E_B-E_A$. The main reason why $V_{AB}$ is so small is because it contains a
multiplicative factor $f$ that can be interpreted as an overlap
factor of the gluon fields mediating the interaction between the quarks.
For the largest squares $f \approx 0.2$. Therefore, in the above discussion,
when comparing A$(d,d)$+FB$(d,d)$ with A$(d,d)$+B$(L-d,d)$, the resultant
small  binding can be interpreted as arising because the colour flux
fields in A have little overlap with those in FB or B. Similarly, when
considering
A$(d,d)$+FA$(d,d)$, the flux field overlaps must be even smaller, since the
flux tubes are ``emitted'' in opposite directions from the quarks
-- as seen in fig.3. The conclusion to
 be drawn from this discussion is that $E_0(7,7)$ seems to be completely
dominated by the
A(7,7)+B(7,7) mixing -- with FA and FB playing minor roles even in the
$16^3 \times 32$ lattice. In other words, even at $r=d=7a$ the effect of
the finite lattice size is unimportant.

\newpage
\vspace{1.0cm}

Table 1. The interquark potential $V_{qq}$, as a function of $(x,y)$.

\begin{itemize}
\begin{enumerate}
\item  The dimensionless quantities $av_{MC}(x/a,y/a)$ correspond to the
Monte Carlo simulation
on the $16^3\times 32$ lattice with $\beta=2.4$ and using
720 measurements from 12 blocks of 240 sweeps each.
$E_0$ and $E_1$ are the ground state and first excited quark-quark potentials.
\item $av_L(x/a,y/a)$ corresponds to $\beta=2.4$ and is from the Coulomb
lattice form in  eqs.(\ref{coul},\ref{coul1})
-- with the additive constant defined such that $v_L(6,0)$ agrees with the
standard Coulomb form at $(x,y)=(6,0)$.

a) Fitting the on-axis $\beta=2.4$ Monte Carlo results.

b) Fitting the on-axis $\beta=2.4$ and 2.5 Monte Carlo results combined.

 \begin{tabular}{|c|c|c|c|} \hline

$(x/a,y/a)$ & $av_{MC}(x/a,y/a)$ & \multicolumn{2}{c|}{$av_L(x/a,y/a)$}  \\
\cline{3-4}
\mbox{}& $aE_0$ \ \ \ \ \ \  $aE_1$ & $\beta=2.4$& $\beta$=2.4 \ {\rm and} \
2.5 \\ \hline
(1,0) & 0.3732(1) \ \ 1.58(2) &0.3539&0.3565 \\
(1,1) & 0.4885(3)  \ \  1.53(5) &0.4816&0.4821 \\
(2,0) & 0.5615(3) \ \  1.51(3)&0.5615&0.5616  \\
(2,1) & 0.6018(4)  \ \ 1.55(2)& 0.6001&0.5999  \\
(2,2) & 0.6680(5) \ \   1.55(2) &  0.6663&0.6659 \\
(3,0) & 0.6797(5)  \ \  1.52(3) & 0.6798&0.6794 \\
(3,1) & 0.6984(6)  \ \  1.54(2) &0.6975&0.6971  \\
(3,2) & 0.741(1)  \ \ 1.56(2) &0.7391&0.7388 \\
(3,3) & 0.796(1) \ \  1.56(4) &0.7939&0.7938  \\
(4,0) & 0.772(1)   \ \ 1.52(4) &0.7721&0.7719  \\
(4,1) & 0.784(1) \ \   1.53(4) &0.7830&0.7829 \\
(4,2) & 0.814(1) \ \  1.57(4)&0.8125&0.8125  \\
(4,3) & 0.856(2) \ \  1.61(4) &0.8551&0.8554 \\
(4,4) &         & 0.9064&0.9070  \\
(5,0) & 0.855(1) \ \  1.58(3) &0.8545&0.8547 \\
(5,1) & 0.864(2) \ \ 1.60(4)&0.8625&0.8628 \\
(5,2) & 0.885(2)  \ \ 1.58(4) &0.8851&0.8856\\
(5,3) & 0.920(3)  \ \ 1.62(5) &0.9197&0.9203 \\
(5,5) &         & 1.0129&1.0143\\
(6,0) & 0.931(2)  \ \ 1.58(5) &0.9322&0.9330  \\
(6,1) & 0.939(2) \ \  1.60(4) &0.9386&0.9394 \\
(6,2) & 0.955(4) \ \  1.62(4) &0.9570&0.9579 \\
(6,3) & 0.984(4) \ \  1.66(5) &0.9858&0.9870 \\
(6,6) &         & 1.1163&1.1186\\
(7,0) &   &1.0073&1.0087  \\
(7,7) &          &1.2180&1.2212  \\
(8,0) &     &  1.0808 &1.0828 \\
(9,0) &     &   1.1532 & 1.1558 \\ \hline
\end{tabular}

\vspace{1.0cm}

\end{enumerate}
\end{itemize}

\vspace{1.0cm}

Table 2.  The energies of four quarks at the corners of squares with sides
$r=d$ in units of $a(\beta=2.4)$ -- see fig.1.
\begin{itemize}
\begin{enumerate}
\item The dimensionless quantities $aE^{16}_i(1600)$ (i=0,1) for 20 blocks

of 80 measurements per block
with the lattice $16^3\times 32$  -- from ref.\cite{GMP2}.
\item $aE^{24}_i(160)$ for 20 blocks of 8 measurements per block with
the lattice $24^3\times 32$.
\end{enumerate}
\end{itemize}

\vspace{0.5cm}

\begin{center}
\begin{tabular}{|c|c|c||c|c|}  \hline
$(\frac{\displaystyle d}{\displaystyle a},\frac{\displaystyle r}
{\displaystyle a})$
& $aE^{16}_0(1600)$ & $aE^{24}_0(160)$ & $aE^{16}_1(1600)
$& $aE^{24}_1(160)$ \\ \hline
(1,1) &--0.0695(5) &--0.0697(3) &0.184(1) &0.182(2)   \\
(2,2) &--0.0582(2) &--0.059(2)&0.142(1) &0.142(2)\\
(3,3) &--0.054(1) &--0.053(1)&0.117(1) &0.114(2) \\
(4,4) & --0.050(1) &--0.049(1) &0.096(1) &0.095(3) \\
(5,5) & --0.047(1) &--0.044(4)&0.072(3) &0.075(5)\\
(6,6) & --0.038(3) &--0.040(3) &0.055(2) &0.056(3) \\
(7,7) & --0.024(5) &--0.05(2)&0.041(5) &0.042(5) \\
\hline
\end{tabular}
\end{center}
\vspace{0.5cm}

\vspace{0.5cm}

Table 3. The energies of four quarks for different values of $\beta$
for $r=d$.

\begin{itemize}
\begin{enumerate}
\item The first three columns are for the Monte Carlo simulation using a

$24^3\times 32$ lattice with $\beta=2.5$.
\item  The second three columns are obtained from the first three by
applying eq.(\ref{scal}) with $\rho=1.437(17)$ and can be compared
with table 2.
\end{enumerate}
\end{itemize}

\vspace{0.5cm}

\begin{tabular}{|c|c|c||c|c|c|}  \hline
$\left(\frac{r}{a(2.5)}\right)$ & $a(2.5)E_0(2.5)$&
$a(2.5)E_1(2.5)$& $\left(\frac{r}{a(2.4)}\right)$ &
$a(2.4)E_0(2.5)$ &$a(2.4)E_1(2.5)$ \\ \hline
1 &--0.0588(1) &0.1608(3)&0.696(8) & --0.084(1) &0.231(3)   \\
2 &--0.0445(3) &0.1172(5) &1.39(2)&--0.064(1) &0.168(2)\\
3 &--0.039(1) &0.096(4) &2.09(2)&--0.056(2) &0.138(6) \\
4 &--0.037(1) &0.091(2) &2.78(3)&--0.053(2) &0.131(3) \\
5 &--0.033(1) &0.081(5)&3.48(4)&--0.047(2) &0.116(7)\\
6 &--0.034(4) &0.078(2) &4.18(5)&--0.049(6) &0.112(3) \\
7 &--0.022(2) &0.072(7)&4.87(6)&--0.032(3) &0.103(10) \\
\hline
\end{tabular}

\vspace{0.5cm}
\vspace{1cm}

Table 4.
 The two lowest eigenvalues for squares and rectangles
as a function of $(d/a,r/a)$ -- in units of the lattice spacing
$a=0.12$ fm.
\begin{itemize}
\begin{enumerate}
\item The dimensionless quantities $aE_0$ and $aE_1$ are from the
Monte Carlo simulation with 1600
measurements on a $16^3\times 32$ lattice and $\beta=2.4$.
\item $aE'_0$ and $aE'_1$ are the two lowest eigenvalues from the model
based solely
on two-quark interactions and expressed as in eqs.(\ref{Ham}--\ref{AVB}).
\item The compromise value of $f=\bar{f}$, which ensures
$E'_i(\bar{f})\approx E_i$, when using eqs.(\ref{Hamf},\ref{NVf}).
\end{enumerate}
\end{itemize}

\begin{center}
\vspace{1cm}

\begin{tabular}{|c|c|c|c|c|c|c|c|}  \hline
$(\frac{\displaystyle d}{\displaystyle a},\frac{\displaystyle r}
{\displaystyle a})$
& $aE_0$ & $aE_1$ & $aE'_0$ & $aE'_1$&$\bar{f}$&$\bar{k}$&$3\bar{\alpha}/4$ \\
\hline
(1,1) &--0.0695(5) &0.184(1) &--0.0769 &0.231&0.886(3)&1.32(4)&1.32(4)   \\
(1,2) &--0.0026(2) &0.505(1)&--0.0039&0.560&0.869(3)&0.61(2)&0.76(2)   \\
(1,3) &--0.0003(3) &0.765(3) &--0.0006 &0.843&0.850(7)&0.35(2)&0.59(3)    \\
(1,4) &0.0003(2) &0.98(1) &--0.0001 &1.078&0.84(2)&0.22(3)&0.46(6)   \\
(2,2) &--0.0582(2) &0.142(1) &--0.0710 &0.213&0.777(3)&0.69(1)&0.69(1) \\
(2,3) &--0.0056(3) &0.324(3) &--0.0122&0.409&0.715(15)&0.56(3)&0.61(4) \\
(2,4) &--0.001(1)&0.52(1)&--0.0035&0.619&0.75(4)&0.31(5)&0.39(7) \\
(3,3) &--0.054(1) &0.117(1) &--0.0775 &0.233& 0.665(4)&0.49(1)&0.49(1) \\
(3,4) &--0.006(1) &0.255(5) &--0.024&0.380&0.57(3)&0.49(5)&0.51(5) \\
(4,4) & --0.050(1) &0.096(1) &--0.090 &0.269&0.519(5)&0.44(1)&0.44(1) \\
(5,5) & --0.047(1) &0.072(3)&--0.106 &0.317&0.36(1)&0.45(1)&0.45(1) \\
(6,6) & --0.038(3) &0.055(1) &--0.123 &0.368&0.259(5)&0.41(1)&0.41(1) \\
(7,7) & --0.024(5) &0.041(5)&--0.140&0.421&0.16(2)&0.41(3)&0.41(3)\\
\hline
\end{tabular}
\end{center}

\vspace{1cm}

\newpage

Table 5. The two lowest energies $(E_{0,1})$ of four colinear
quark systems --  in units of $a(\beta =2.4)$.

1) 20 blocks of 320 sweeps i.e. 1600 measurements

2) The dimensionless quantities $aE_i$ are from the Monte Carlo simulation
and the $aE'_i$ from the model.

\begin{center}
\vspace{0.2cm}

\begin{tabular}{|c|c|c|c|c|} \hline
$(\frac{\displaystyle r}{\displaystyle a},\frac{\displaystyle d}
{\displaystyle a})$
& $aE_0$ & $aE_1$ &  $aE'_0(f=1)$ & $aE'_1(f=1)$    \\ \hline
(2,1) &--0.0026(3) &0.418(3) &--0.00357   &0.459         \\
(3,1) &--0.0002(3) &0.736(7)&--0.00028   &0.800          \\
(4,1) &  0.0002(3) &0.99(2) &--0.00003   &1.057          \\
(3,2) &--0.019(2) &0.204(3) &--0.0231    &0.250         \\
(4,2) &--0.002(3) &0.473(4) &--0.00161  &0.530        \\
(5,2) & 0.000(1) &0.69(1) &--0.00021  &0.767         \\
(4,3) &--0.039(3) &0.142(3) &--0.0482   &0.185          \\
(5,3) & --0.003(3) &0.377(5) & --0.0035  &0.425           \\
(6,3) & --0.005(5) &0.58(3)& --0.00051 &0.653           \\
\hline
\end{tabular}
\end{center}
\newpage
\vspace{0.2cm}
Table 6. The comparison of the four colinear quark energies $E_{0,1}$
of table 5 with the model of section 4.
\begin{itemize}
\begin{enumerate}
\item The two lowest model energies
$aE'_i(\bar{f})$ using the compromise value of $\bar{f}$.
\item The compromise value of $\bar{k}$ from
$\bar{f}=\exp\left(-\bar{k}b_s<r_{ij}^2>\right)$

with $b_Sa_{2.4}^2=0.0691$ from section 2.
\item The compromise value of $\bar{p}$ from
$\bar{f}=\exp\left(-\bar{p}\sqrt{b_s}L\right)$.
\end{enumerate}
\end{itemize}

\begin{center}

\vspace{0.2cm}

\begin{tabular}{|c|c|c|c|c|c|} \hline
$(\frac{\displaystyle r}{\displaystyle a},\frac{\displaystyle d}
{\displaystyle a})$
&$aE'_0(\bar{f})$ & $aE'_1(\bar{f})$& $\bar{f}$&$\bar{k}$&$\bar{p}$ \\ \hline
(2,1) &--0.00285(7) &0.418(3)& 0.88(1)&0.54(6)&0.079(8)   \\
(3,1) &--0.00021(1)&0.736(7)&0.87(2) &0.31(5)&0.067(9)  \\
(4,1) &--0.000025(2) &0.99(2)&0.89(4)&0.14(5)&0.042(15) \\
(3,2) &--0.0186(3)&0.204(3)&0.85(1)&0.27(2)&0.063(6) \\
(4,2) &--0.00116(4) &0.473(4)&0.84(1)&0.19(2)&0.055(5) \\
(5,2) &--0.00015(1) &0.69(1)&0.83(3)&0.14(3)&0.051(9) \\
(4,3) &--0.0424(5)&0.141(3)&0.86(1)&0.14(1)&0.043(4) \\
(5,3) &--0.0026(1) &0.377(5)&0.85(2)&0.10(1)&0.039(5) \\
(6,3) &--0.0003(1) &0.58(3)&0.8(1)&0.09(5)&0.04(3) \\
\hline
\end{tabular}
\end{center}

\newpage
\vspace{1cm}

Table 7. The values of  $f_{AB},f_{AC}$ and $f_{BC}$
satisfying  $E'_0(f_{AB})=E_0 \ ,E'_0(f_{AC})=E_0$ and $E'_0(f_{BC})=E_0$
and also, in the brackets [...], the values of $f$ to
ensure $E'_1(f)=E_1$ .

\begin{center}
\begin{tabular}{|c|c|c|c|} \hline
$(\frac{\displaystyle r}{\displaystyle a},\frac{\displaystyle d}
{\displaystyle a})$
&$f_{AB}(E$) &$f_{AC}(E$)  & $f_{BC}(E$)  \\ \hline
(2,1) &0.84(5) &0.86(5)& 0.999(1)   \\
      &[0.88(1)]&[0.76(3)]& [0.51(3)]  \\
(3,1) &0.8(6)&0.8(6)&1.000(1)  \\
      &[0.87(2)]& [0.85(2)]& [0.58(4)] \\
(3,2) &0.86(7)&0.92(4)&0.99(1) \\
      &[0.85(1)]&[no fit]& [no fit] \\
(4,2) &1.1(8) &1.1(6)&1.00(1) \\
      &[0.84(1)]&[0.74(3)] & [0.46(3)]\\
(4,3) &0.78(7)&0.92(3)&0.95(2) \\
      & [0.86(1)]&[ no fit]& [no fit] \\
(5,3) &0.9(6) &0.9(6)&0.99(1) \\
      & [0.85(2)]& [0.66(6)]& [0.40(5)] \\
\hline
\end{tabular}
\end{center}

\vspace{1cm}

\newpage
{\bf \ \ \ \  Figure Captions}

\vskip 0.5cm

Fig.1 a) The basic four-quark configuration.

\hspace*{1.0cm} b) The meson-like partition
$A=M_{13}M_{24}=[q_1q_3]^0[q_2q_4]^0$.

\hspace*{1.0cm} c) The meson-like partition
$B=M_{14}M_{23}=[q_1q_4]^0[q_2q_3]^0$.

\vskip 0.5cm

Fig.2 Check on the finite lattice size.
\begin{itemize}
\begin{enumerate}
\item Solid line -- The Monte Carlo simulation on the $16^3\times 32$ lattice
with $\beta=2.4$.
\item Cross points -- The Monte Carlo simulation on the $24^3\times 32$
lattice  -- for clarity the results at $r$ are plotted at $r+0.25a$.

\end{enumerate}
\end{itemize}

  \vspace{0.5cm}
Fig.3 Configurations contributing to the finite size correction.

\indent a) State A  \ \  b) State F

\vspace{0.5cm}

Fig.4. Check on scaling.
\begin{itemize}
\begin{enumerate}
\item Solid line -- The Monte Carlo simulation on the $16^3\times 32$ lattice
with $\beta=2.4$ -- as in fig.2.
\item Cross points -- The Monte Carlo simulation on the $24^3\times 32$
lattice with $\beta=2.5$ and converted
into the corresponding $\beta=2.4$ values using eq.(\ref{scal}) and
$\rho=1.437(17).$
\end{enumerate}
\end{itemize}

\vspace{0.5cm}

Fig. 5  Values of $\rho(k)$ for different lattices and values of $\beta$.

\begin{itemize}
\begin{enumerate}

\item $E_0(r=d=2a)$ for  $\beta=2.4$ on the $16^3\times 32$ lattice
with the number of measurements $N_M=10$ for a) the number of blocks $N_B=10$
and b) $N_B=40$, and c) $N_M=40$ and $N_B=20$ d) $N_M=20$ and $N_B=40$.
\item $\beta=2.5$ on the $24^3\times 32$ lattice with $N_M=30$ and $N_B=22$,
and e) for the energy $E_0(r=d=2a)$,  f) for the energy $E_1(r=d=2a)$.

\end{enumerate}
\end{itemize}

\vskip 0.5cm

Fig.6 Four quarks in a straight line. The three configurations A,B and C.

\vskip 0.5cm

Fig.7 The values of $\bar{f}$,$\bar{k}$ and $\bar{p}$ from table 6 -- as
a function of $(r/a,d/a)$

\vspace{0.5cm}

 Fig. 8  The two flux paths between quarks at $x_1$ and $x_2$.

\hspace*{1.0cm} a) The direct path $A$

\hspace*{1.0cm} b) The path $FA$ that encircles the spatial periodic boundary.

\noindent All $U$ links between $x_0$ and $x_0+a$ are multiplied by an
element (z) of the group Z(N).

\end{document}